\documentstyle[aps,epsf]{revtex}

\begin{document}
\title{Ground state of antiferromagnetic ordering in fullerene $C_{60}$ molecule}
\author{C.V. Usenko$^1$}
\author{V.C. Usenko$^2$}
\address{$^1$ Taras Shevchenko Kiev University, Physical Department. 6, Academician\\
Glushkov pr., Kiev 03127, Ukraine }
\address{$^2$ Institute of Physics, NAS of Ukraine. 46, Nauky pr., Kiev 03022, Ukraine}
\date{\today}
\maketitle

\begin{abstract}
Theoretical study of mutual orientation of fullerene $C_{60}$ molecule atom
spins is presented in this work. Spin-spin interaction was described by
Habbard's model. Existence of antiferromagnetic sturcture of spin sub-system
in ground state is found. It is also shown that spin distribution
practically does not depend on weak anisotropy due to Jan-Teller effect.
\end{abstract}

\section{INTRODUCTION}

At theoretical study of a fullerene molecule $C_{60}$ , general approach is
to use the localized electrons model \cite{r3}\cite{r7}. Though question on
if in fullerene there exists an ordered spin subsystem was practically not
considered. In general, study of magnetic properties of a fullerene molecule 
$C_{60}$ is now at the very beginning. Theoretical works by Gunnarson et
al., in particular, ''Mott-Habbard insulators for systems with orbital
degeneracy'' \cite{r1}, where application of Habbard model to simplest
lattices with space topology close enough to topology of fullerens, in
particular, tetragon and triangle, is considered, are known. Authors of that
work have practically approached the problem of study of doped fullerenes
magnetic properties. Recently an experimental work by Schilder et al.\cite
{r6}, ''The role of TDAE for the magnetism in $C_{60}$ '', appeared as well,
where magnetic properties of a doped fullerene are studied, in particular -
effect of admixture on those properties. In this work theoretical aspects of
fullerene $C_{60}$ magnetic properties are studied, it is proved that there
exists a ground state of fullerene molecule spin subsystem; properties of
the ground state are proved to be practically independent from the binding
in fullerene molecule anisotropy.

In the case of anti-ferromagnetism all or some of magnetic moments of atoms
are oriented in such a way (usually those are anti-parallel) that total
magnetic moment of crystal unit cell is zero (or is a small part of moment
of an atom). Here anti-ferromagnetism differs from ferromagnetism, in the
case of which same orientation of all atomic magnetic moments leads to high
magnetization of the body. Responsible for anti-ferromagnetism coming to
existence is exchange interaction, that is trying to arrange spins (and,
thus magnetic moments) in anti-parallel way; in such a case exchange
integral has a negative value. Type of magnetic ordering is determined by
magnetic atomic structure, symmetry of which is described by point and space
groups of magnetic symmetry. Unit cell of magnetic structure may coincide
with crystallographic one, or it can be multiple of it (for instant, have a
twice larger period). Magnetic structure of anti-ferromagnets is described
by a set of embedded magnetic sub-lattices, each with magnetization $\vec{M}%
_{j}$ . In all anti-ferromagnets, except of those with weak ferromagnetism,
under absence of external magnetic field, $\sum\limits_{j}{\vec{M}_{j}}=0$ .
In the case of magnetic sub-lattices application one considers for
anti-ferromagnets properties description (in particular, to describe
transition from paramagnetic state to anti-ferromagnetic one) following
parameters: vector of anti-ferromagnetism $\vec{L}=\vec{M}_{1}-\vec{M}_{2}$
and vector of magnetization $\vec{M}=\vec{M}_{1}+\vec{M}_{2}$ \ (for two
sub-lattices).

The term ''magnetic sub-lattice'' came into existence when number of
sub-lattices for known crystals did not exceed two, i. e. when for magnetic
structures of crystals only the simplest ones were known: ferromagnetic
structure (one sub-lattice) and collinear anti-ferromagnetic (two
sub-lattices, later on Neel's collinear ferromagnetic structure being added,
having in simplest cases two sub-lattices as well). Later on magnetics with
more complicated, non-collinear magnetic structures have been found, having,
respectively, a larger number of sub-lattices (three, four and more).
Obviously, number of sub-lattices can not exceed number of atoms in a
magnetic cell. Properties of ferromagnets with multiple sub-lattices, in
particular, effects like sub-lattices overturn (jump turn of atomic magnetic
moments at external magnetic field, applied along the axis of symmetry,
coming to some critical value, related to anisotropy of magnetic
susceptibility), relativistic distortion of sub-lattices mutual orientation,
are studied in works by Marchenko et al.\cite{r3,r4}.

In wide-spread carbon compounds, like graphite and diamond, it is not
possible to isolate the spin state of one atom, because of collectivization
of valence electrons (this results from strong interatomic interaction and
crystalline structure of such substances). One of consequences is that those
matters do not show magnetic properties. Another situation one finds for the
fullerene molecule $C_{60}$ , where number of atoms is much less than in the
case of diamond or graphite and this makes it possible to suppose all
electrons to be localized on atoms. For molecules of such a type one should
expect existence of some order in atomic spins orientation, and this can
influence magnetic properties. Since interaction of p-electrons spins seeks
to turn spins of neighbouring atoms in opposite directions,
anti-ferromagnetism should be specific for $C_{60}$ molecules.

In accordance to Hund's rule, the lowest energy is specific for the term
with the maximal possible at given electronic configuration value of the
total spin of electrons S and the largest (possible at that S) value of
total orbital moment L. Thus, if spin of an atom (being, really, the sum of
spins of electrons in outer shell) is non-zero, it, in accordance to Hund's
rule, takes the maximal value. Average spin of an atom is a part of the
total moment of momentum of the valence electrons system, and its value
depends on electronic configuration. For a separate carbon atom
configuration $s^{2}p^{2}$ is specific, where spins of $s$-electrons have
opposite orientation, though for the same atom in chemical compounds, in
fullerene among those, configuration $sp^{3}$ is more often, where all
electron spins, in accordance to Hund's rule, are parallel, and total atomic
spin is to equal 2.

Intrinsic magnetic moment of an atom, thus moment without external field,
is: 
\begin{equation}
\vec{\mu}_{o}=-\mu _{B}(\vec{L}+2\vec{S})  \label{e1}
\end{equation}
where $\mu _{B}$ is the Bohr magneton, $\vec{L}$ - total orbital moment of
an atom, $\vec{S}$ - total spin of an atom. In a carbon atom of a fullerene
molecule orbital moments of all electrons are compensated, thus total
orbital moment is $\vec{L}=0$ . Hence, intrinsic magnetic moment of each
atom is collinear to its spin vector. Thus, to disclose carbon atoms in
molecule $C_{60}$ configuration it is enough to determine configuration of
atomic spins of the molecule, thus the spin (or, this is the same,
anti-ferromagnetic) ordering; question on localization and configuration of
spins is, hence, closely related to magnetic properties of a molecule. In
the study of spin ordering account of carbon atoms spin - spin interaction,
determining, in fact, configuration, is needed. Main state of ordering is in
such a case the one with minimal energy of spin interaction.

\section{PROBLEM FORMULATION}

Neglecting spin-spin interaction, one can consider in zero approximation
spins of different atoms to be independent, and in first-order approximation
interaction of those is determined by Heizenberg's Hamiltonian, 
\begin{equation}
H_{H}={\frac{1}{2}}\sum\limits_{i\neq j}{J_{i,j}\hat{\vec{S}}_{i}\cdot \hat{%
\vec{S}}_{j}}  \label{e21}
\end{equation}
Interaction of spins is determined by the overlap integrals for coordinate
wave functions of valence electrons 
\begin{equation}
J_{i,j}=\left\langle \varphi _{i}|\varphi _{j}\right\rangle  \label{e22}
\end{equation}
In first-order approximation one can take into account interaction of first
neighbors only, neglecting wave functions overlapping for all other pairs of
atoms. In $C_{60}$ molecule each atom has three nearest neighbors, similar
to diamond or graphite, thus in interaction energy inputs of only three
nearest atoms are left: 
\begin{equation}
\left\langle H_{H}\right\rangle =\frac{1}{2}\sum_{i}\left( J_{1}\left\langle 
\overrightarrow{S}_{i}\right\rangle \cdot \left\langle \overrightarrow{S}%
_{i1}\right\rangle +J_{2}\left\langle \overrightarrow{S}_{i}\right\rangle
\cdot \left\langle \overrightarrow{S}_{i2}\right\rangle +J_{3}\left\langle 
\overrightarrow{S}_{i}\right\rangle \cdot \left\langle \overrightarrow{S}%
_{i3}\right\rangle \right)  \label{e23a}
\end{equation}
It is taken into account here that because of symmetry overlap integrals
with respective neighbor for each atom coincide, and indices $i1,i2,i3$
stand for number of respective neighboring atom. Jan-Teller effect removes,
in a part, degeneration of values for overlap integrals, though difference
in distances between atoms (1.45, 1.45 and 1.40) is small enough to neglect
it in model approximation. 
\begin{equation}
\left\langle H_{H}\right\rangle =\frac{1}{2}J\sum_{i}\left( \left\langle 
\overrightarrow{S}_{i}\right\rangle \cdot \left\langle \overrightarrow{S}%
_{i1}\right\rangle +\left\langle \overrightarrow{S}_{i}\right\rangle \cdot
\left\langle \overrightarrow{S}_{i2}\right\rangle +\left\langle 
\overrightarrow{S}_{i}\right\rangle \cdot \left\langle \overrightarrow{S}%
_{i3}\right\rangle \right)  \label{e23b}
\end{equation}
Sign of overlap integrals defines the properties - either ferromagnetic, or
paramagnetic - to take place. For carbon overlap integral is positive, thus
parallel spins would correspond to maximum in the spin-spin interaction
energy, while minimal energy is to be observed for absolutely antiparallel
spins. As it was mentioned above, surface of the $C_{60}$ molecule is
arranged of pentagons and hexagons, with carbon atoms in vertices. Before
considering molecule as a whole we find configurations of atomic spins for
atoms in vertices of two such polygons, under condition of antiparallel
orientation of spins (this condition follows from requirement for minimum of
energy of interaction between spins). In the case of hexagon there are no
problems with arrangement - antiparallel orientation for each two nearest
neighbors completely satisfies this criterion. Though in case of pentagon
situation is more complicated. Because of number of vertices being in this
case uneven, it is not possible to arrange for such a figure a really
antiparallel system of spins, orthogonal to its plane. Thus it turns out
that it is not possible to arrange for a fullerene a usual
anti-ferromagnetic ordering, with two magnetic lattices; in reality more are
needed. These considerations make evident that orientation of spins in a
fullerene molecule shall be of such a type that makes energy of spin-spin
interaction minimal for the molecule as a whole only, and not for each pair
of neighboring atoms. Such a minimum for a molecule as a whole is possible
in the case of long-range atomic spins ordering only, i. e. under condition
of totally correlated spin directions for not only nearest neighbors, but
all atoms in the molecule as well. Aim of this work is to find such an
arrangement.

Problem on atomic spins interaction for a $C_{60}$ molecule in the model of
localized electrons is studied under following assumptions:

\begin{itemize}
\item  Only spin-spin interaction for nearest neighbors (three of those for
each atom) is taken into account;

\item  Constants of interaction are considered as being same for each pair
of atoms (Jan-Teller effect is neglected);

\item  Possible changes in average values of spins for each atom are
neglected (semi-classical approach).
\end{itemize}

Complexity of the problem in comparison to anti-ferromagnetic crystals
models is in absence of long-range ordering, leading to necessity of
studying spins for all 60 atoms simultaneously. Space orientation of spins
is determined by $2\times 60$ angles, or $3\times 60$ unit vector
components. Since symmetry group of $C_{60}$ molecule is simple, possibility
to isolate sub-lattices with same arrangement of spins does not exist.
Specific structure of neighbors arrangement for each atom does not allow to
use same notation for the system of neighbors, and makes it again necessary
to study spins of all atoms. With account of all approximations, one can
represent the spin-spin energy of interaction in $C_{60}$ molecule as 
\begin{equation}
E=\frac{{JS^{2}}}{2}\sum\limits_{i=1..60}{\sum%
\limits_{k_{i}=1_{i},2_{i},3_{i}}{\vec{n}_{i}\cdot \vec{n}_{k_{i}}}}
\label{e24}
\end{equation}
Here it as taken into account that spin of each atom can vary in direction
only, thus for the ith atom ''i'' spin we use notation 
\begin{equation}
\vec{S}_{i}=S\vec{n}_{i},  \label{e25}
\end{equation}
where vector of direction of the spin $\vec{n}_{i}$ is a unit vector, and
indices $1_{i},2_{i},3_{i}$ stand for neighbors of ith atom . Thus problem
is in finding the minimum of spin-spin interaction energy (\ref{e24}) under
condition of preserving lengths of all vectors $\vec{n}_{i}$, which imposes
an additional condition 
\begin{equation}
\left| {\vec{n}_{i}}\right| ^{2}=1;i=1..60.  \label{e27}
\end{equation}
It is easy enough to determine the lower limit of minimal energy: for each
separate node minimal energy would be observed in the case of spins of all
neighboring atoms being arranged in parallel to each other and in
antiparallel orientation with respect to spin of that node. In such a case
each scalar product for that node is -1, and if it was possible to arrange
all the spins in such a way, total energy would be 
\begin{equation}
E_{hyp.\min }=-90JS^{2}.  \label{e28}
\end{equation}
In reality, because of specific topology, it is not possible to arrange all
spins of the $C_{60}$ molecule in antiparallel way, thus energy of spin-spin
interaction, even for ground state, exceeds this value.

Consider first a much simpler, as from point of view of number of atoms,
though a topologically similar imaginary molecule, formed by 4 atoms,
situated in vertices of a regular tetrahedron. Spin-spin interaction energy
can be represented in explicit form: 
\begin{equation}
E_{ss}=JS^{2}\left( {\vec{n}_{1}\cdot \vec{n}_{2}+\vec{n}_{1}\cdot \vec{n}%
_{3}+\vec{n}_{1}\cdot \vec{n}_{4}+\vec{n}_{2}\cdot \vec{n}_{3}+\vec{n}%
_{2}\cdot \vec{n}_{4}+\vec{n}_{3}\cdot \vec{n}_{4}}\right) .  \label{e29}
\end{equation}
If each couple of spins was antiparallel, energy of such a state would be 
\begin{equation}
E_{hyp.\min }=-6JS^{2}.  \label{e210}
\end{equation}
Symmetry considerations make it possible to find out that minimum is
observed with all the spins oriented in direction from the center of
tetrahedron, or towards its center. Each scalar product equals $-\frac{1}{3}$%
, and sum of those -$E_{ss}=-2JS^{2}$, this is three times larger than the
lower limit estimation mentioned above.

This example makes evident that there is a conventional possibility to
consider the state with minimal energy as being equilibrium only.
Substantial peculiarity of tetrahedron model is in fact that angles formed
by directions from the molecule center to each node exceed in this case the
right ones, thus even in the most symmetric state main energy, falling for
each state, is negative. Even for the next in complexity model, a cube, such
a radial spins arrangement can provide not more than a zero value of energy
of interaction with nearest neighbors. For more complicated molecules, with $%
C_{60}$ among those, angles between spins at radial arrangement tend to
zero, thus energy of spin-spin interaction for such an arrangement shall
tend to absolute maximum. In fact, this means that if for a tetrahedron
radial arrangement of spins is optimal (thus it corresponds to minimal
energy), similar radial arrangement in the case of fullerene shall lead to
opposite result.

\section{SPIN SUB-SYSTEM GROUND STATE}

Equations of motion for the spin system have steady-state solutions only in
the case of sum of spins of nearest neighbors of each atom having parallel
to that atom's spin orientation. Really, if one rewrites the system of
equations of motion as follows: 
\begin{equation}
\frac{{d\vec{n}_{i}}}{{dt}}=JS^{2}\vec{n}_{i}\times \sum\limits_{j\neq i}{%
\vec{n}_{j},}  \label{e31}
\end{equation}
it is easy to see that under condition 
\begin{equation}
\sum\limits_{k_{i}}{\vec{n}_{k_{i}}}=\lambda _{i}\vec{n}_{i}  \label{e32}
\end{equation}
all spins are in rest, and energy of spin-spin interaction equals 
\begin{equation}
E_{ss}=JS^{2}\sum\limits_{i=1..60}{\lambda _{i}.}  \label{e33}
\end{equation}
The lowest of all possible values of the sum of parameters $\lambda _{i}$
corresponds to the state with minimal energy. With account of symmetry of
the molecule it is possible to consider $\lambda _{i}$ as being the same for
all the nodes. Besides that, all vectors $\vec{n}_{i}$ are to be of the same
(unit) length, thus one has to add to the system (\ref{e32}) conditions
being imposed on the vector lengths 
\begin{equation}
\left| {\vec{n}_{i}}\right| ^{2}=1;i=1..60.  \label{e34}
\end{equation}
Those conditions make the problem to be non-linear.

System of equations (\ref{e32}) can be divided into three similar systems 
\begin{equation}
\begin{array}{l}
\sum\limits_{k_{i}}{n_{xk_{i}}}=\lambda _{i}n_{xi} \\ 
\sum\limits_{k_{i}}{n_{yk_{i}}}=\lambda _{i}n_{yi} \\ 
\sum\limits_{k_{i}}{n_{zk_{i}}}=\lambda _{i}n_{zi}
\end{array}
.  \label{e35}
\end{equation}
Input data are coordinates of atoms of fullerene molecules. All calculations
have been done in the program package Maple V.5 by means of a system of
analytic calculations, being a part of the package. Each step of solving the
problem was implemented as a separate program. To solve the problem
following algorithm is used:

\begin{enumerate}
\item  Atoms of the fullerene molecule are sequentially numbered. For each
atom numbers of three nearest neighbors are found. Next, a matrix $60\times
60$ is arranged, where in each line there are values ''1'' in elements,
numbers of which coincide with numbers of nearest neighbors of the atom
corresponding to number of line, all other elements have values ''0''. It is
obvious that since approximation of three nearest neighbors is considered,
in different positions of each line there are three values ''1''. This
matrix (let it be L) is the matrix of the system of equations (\ref{e35}).

\item  Next, eigenvalues and eigenvectors of matrix L are found. Minimum
eigenvalue $\lambda _{_{\min }}$, corresponding to the state of spin
subsystem with minimal ordering energy, is of triple degeneracy. As one
should expect, respective eigenvectors (60 components each) form the set of
coordinates $\vec{n}_{i}$ for the problem.

\item  While the study it turned out that eigenvectors found are not
orthonormal ones, i. e. coordinates of vectors $\vec{n}_{i}$ were found not
in orthogonal basis. Orthonormalization of eigenvectors was performed.

\item  By components of orthonormal eigenvectors of matrix L vectors $\vec{n}%
_{i}$ of spin directions for each node are constructed. Checkup of vector
lengths congruence is performed.

\item  By respective minimal eigenvalue of matrix $L$ energy of ground state
of the spin sub-system ordering is found. Last, graphic representation of a
fullerene molecule and atomic spins vectors in eigenvalue is performed.
\end{enumerate}

\section{ RESULTS}

The minimum eigenvalue found 
\[
\lambda _{0}\text{={\rm -2}{\rm .6180,}}
\]
as it is mentioned above, has triple degeneration, and a set of respective
orthonormal eigenvalues that satisfies conditions of normalization is found,
thus the problem on the ground state of spin sub-system has been solved.
Energy corresponding to minimal eigenvalue is 
\[
E_{ss}=60JS^{2}\lambda _{0},
\]
this is a little (by 13\%) higher than minimum. Following picture \ref{fig}
shows the $C_{60}$ molecule, with directions of spins of all the atoms.

\begin{figure}[tbp]
\caption{Directions of spins.}
\label{fig}
\end{figure}

\section{CONCLUSIONS}

Results obtained in this study make evidence of stationary states for the
fullerene molecule spin subsystem ordering existence in localized spin
model. Value of energy for the spin sub-system ground state is a bit higher
than theoretically possible minimum. The spin sub-system spectrum
practically does not depend on weak anisotropy due to Jan-Teller effect.

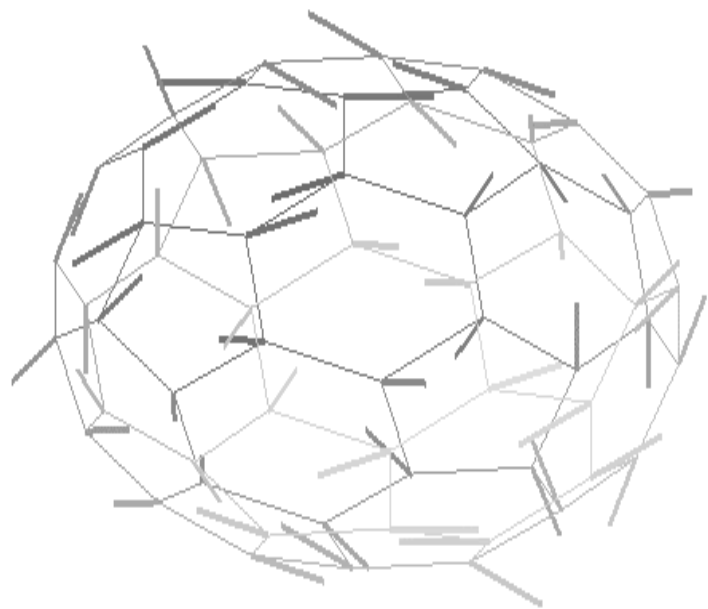
\end{document}